\documentclass[12pt,prd,aps,amssymb,amsmath,tightenlines,showpacs,a4paper]{revtex4}
\begin{document}

\def\half{\textstyle{\frac{1}{2}}}
\def\cP{\mathcal P}
\def\cC{\mathcal C}
\def\cT{\mathcal T}

\rightline{preprint LA-UR-07-0431}

\title{Spontaneous Breaking of Classical $\cP\cT$ Symmetry}

\author{Carl~M.~Bender${}^1$\footnote{Permanent address:
Department of Physics, Washington University, St. Louis MO 63130, USA}}
\email{cmb@wustl.edu}

\author{Daniel~W.~Darg${}^2$}\email{daniel.darg@imperial.ac.uk}

\affiliation{${}^1$Center for Nonlinear Studies, Los Alamos National Laboratory,
Los Alamos, NM 87545, USA}

\affiliation{${}^2$Blackett Laboratory, Imperial College, London SW7 2BZ, UK}

\begin{abstract}
The classical trajectories of the family of complex $\cP\cT$-symmetric
Hamiltonians $H=p^2+x^2(ix)^\epsilon$ ($\epsilon\geq0$) form closed orbits. All
such complex orbits that have been studied in the past are $\cP\cT$ symmetric
(left-right symmetric). The periods of these orbits exhibit an unusual
dependence on the parameter $\epsilon$. There are regions in $\epsilon$ of
smooth behavior interspersed with regions of rapid variation. It is demonstrated
that the onset of rapid variation is associated with strange new kinds of
classical trajectories that have never been seen previously. These rare kinds of
trajectories are {\it not} $\cP\cT$ symmetric and occur only for special
rational values of $\epsilon$.
\end{abstract}
\pacs{11.30.Er, 45.50.Dd, 02.30.Oz}
\maketitle

\section{Introduction}
\label{s1}
The classical trajectories $x(t)$ of the family of complex $\cP\cT$-symmetric
Hamiltonians
\begin{equation}
H=p^2+x^2(ix)^\epsilon\qquad(\epsilon\geq0)
\label{e1}
\end{equation}
have been examined in detail \cite{r1}. One can plot graphs of these
trajectories by solving numerically the system of Hamilton's differential
equations \cite{dargweb}
\begin{equation}
{\dot x}=\frac{\partial H}{\partial p}=2p,\quad
{\dot p}=-\frac{\partial H}{\partial x}=-(2+\epsilon)x(ix)^\epsilon
\label{e2}
\end{equation}
for a given set of initial conditions $x(0)$, $p(0)$. Since $x(0)$ and $p(0)$
are not necessarily real numbers and the differential equations (\ref{e2}) are
complex, the classical trajectories are curves in the complex-$x$ plane. It is
known \cite{r1} that for $\epsilon\geq0$ nearly all trajectories are closed
curves. (When $\epsilon$ is a positive integer, it is possible for trajectories
that originate at some of the turning points to run off to infinity. However, we
are not interested here in these isolated singular cases. When $\epsilon<0$, all
trajectories are open curves.) If $\epsilon$ is noninteger, there is a branch
cut in the complex-$x$ plane, and we take this cut to run from $0$ to $\infty$
along the positive-imaginary axis. Thus, it is possible for a closed classical
trajectory to visit many sheets of the Riemann surface before returning to its
starting point.

The non-Hermitian Hamiltonians (\ref{e1}) are remarkable because when they are
quantized, their spectra are entirely real and positive \cite{r2,r3}. Moreover,
these Hamiltonians specify a unitary time evolution \cite{r4} of the vectors in
the associated Hilbert space. Thus, it is important to understand the nature of 
the complex classical systems underlying these quantum systems.

Several studies \cite{r5,r6} of the classical trajectories $x(t)$ of complex
Hamiltonians were done prior to the work in Ref.~\cite{r1} and from all these
studies many features of the complex trajectories of $\cP\cT$-symmetric
Hamiltonians (\ref{e1}) are known. However, some of the conclusions of the
earlier work are wrong. For example, when $\epsilon\geq0$, the $\cP\cT$ symmetry
of the quantum-mechanical theory is unbroken \cite{r4}, and based on these
studies it was believed that all classical orbits are $\cP\cT$ symmetric. [We
say that an orbit is $\cP\cT$ {\it symmetric} if the orbit remains unchanged
upon replacing $x(t)$ by $-x^*(-t)$. Such an orbit has mirror symmetry under
reflection about the imaginary axis on the principal sheet of the Riemann
surface.] While the equations of motion (\ref{e2}) exhibit $\cP\cT$ symmetry, it
is not required that the solutions to these equations also exhibit $\cP
\cT$ symmetry, but in all previous numerical studies only $\cP\cT$-symmetric
orbits were found.  We will show in this paper that there are also rare
trajectories that are {\it not} $\cP\cT$ symmetric, and we will argue that these
new kinds of orbits explain the strange fine-structure behavior of the periods
of the orbits that was first reported in Ref.~\cite{r1}.

This paper is organized as follows: In Sec.~\ref{s2} we review briefly the
earlier work on classical trajectories. Then, in Sec.~\ref{s3} we present new
findings that help us to grasp the underlying reasons for the appearance of the
elaborate and intricate structures of the classical trajectories that were
described in Ref.~\cite{r1}. In Sec.~\ref{s4} we make some concluding remarks.

\section{Brief summary of previous numerical studies}
\label{s2}

To construct the classical trajectories, we note that the Hamiltonian in
(\ref{e1}) is a constant of the motion. This constant (the energy $E$) may be
chosen to be 1 because if $E$ were not 1, we could rescale $x$ and $t$ to make
$E=1$. Because $p(t)$ is the time derivative of $\half x(t)$ [see (\ref{e2})],
the trajectory $x(t)$ satisfies a first-order differential equation whose
solution is determined by the initial condition $x(0)$ and the sign of $\dot x(0
)$.

The simplest version of the Hamiltonian (\ref{e1}) is the harmonic oscillator,
which is obtained by setting $\epsilon=0$. For the harmonic oscillator the
turning points, which are the solutions to $x^2=1$, lie at $x=\pm1$. If we chose
$x(0)$ to lie between these turning points, then the classical trajectory
oscillates between the turning points with period $\pi$. However, while the
harmonic-oscillator Hamiltonian is real, it still has complex classical
trajectories. To generate one of these trajectories, we choose a value for $x(0
)$ that does not lie between $\pm1$ and find that the resulting trajectories are
ellipses in the complex plane \cite{r1}. The foci of these ellipses are the
turning points at $x=\pm1$ \cite{r1}. The period for all of these closed orbits
is $\pi$. The constancy of the period is due to the Cauchy integral theorem
applied to the path integral that represents the period. The (closed) contour of
integration encircles the square-root branch cut that joins the turning points.

\begin{figure*}[t!]
\vspace{5.0in}
\includegraphics{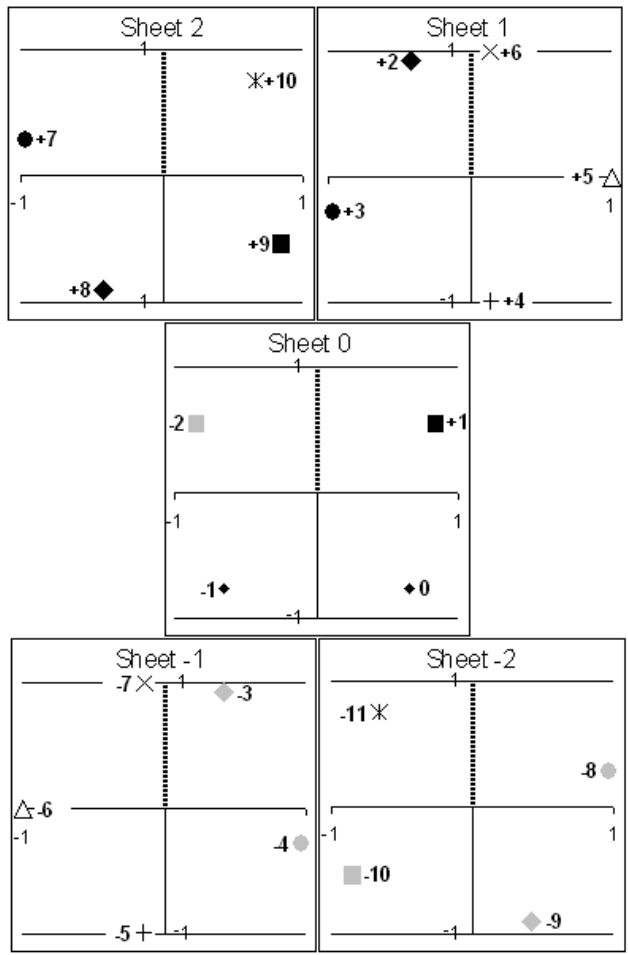}
\caption{Locations of the turning points for $\epsilon=\frac{12}{5}$. There are
11 $\cP\cT$-symmetric pairs of turning points, with each pair being mirror
images under reflection through the imaginary-$x$ axis on the principal sheet.
All 22 turning points lie on the unit circle on a five-sheeted Riemann surface,
where the sheets are joined by cuts on the positive-imaginary axis.}
\label{fig1}
\end{figure*}

As $\epsilon$ increases from 0, the turning points at $x=1$ (and at $x=-1$)
rotate downward and clockwise (anticlockwise) into the complex-$x$ plane. These
turning points are solutions to the equation $1+(ix)^{2+\epsilon}=0$. When
$\epsilon\neq2$, this equation has many solutions that all lie on the unit
circle and have the form
\begin{equation}
x=\exp\left(i\pi\frac{4N-\epsilon}{4+2\epsilon}\right)\quad(N~{\rm integer}).
\label{e3}
\end{equation}
(This notation differs slightly from that used in Ref.~\cite{r1}.) These turning
points occur in $\cP\cT$-symmetric pairs (pairs that are symmetric when
reflected through the imaginary axis) corresponding to the $N$ values $(N=-1,~N=
0)$, $(N=-2,~N=1)$, $(N=-3,~N=2)$, $(N=-4,~N=3)$, and so on. We label these
pairs by the integer $K$ ($K=0,~1,~2,~3,~\ldots$) so that the $K$th pair
corresponds to $(N=-K-1,~N=K)$. The pair of turning points on the real-$x$ axis
for $\epsilon=0$ deforms continuously into the $K=0$ pair of turning points when
$\epsilon>0$. When $\epsilon$ is rational, there are a finite number of turning
points in the complex-$x$ Riemann surface. For example, when $\epsilon=\frac{12}
{5}$, there are 5 sheets in the Riemann surface and 11 pairs of turning points.
The $K=0$ pair of turning points are labeled $N=-1$ and $N=0$, the $K=1$ pair
are labeled $N=-2$ and $N=1$, and so on. The last ($K=10$) pair of turning
points are labeled $N=-11$ and $N=10$. These turning points are shown in
Fig.~\ref{fig1}.

As $\epsilon$ increases from 0, the elliptical complex trajectories for the
Harmonic oscillator begin to distort. However, the trajectories remain closed
and periodic except for special singular trajectories that run off to complex
infinity. These singular trajectories only occur when $\epsilon$ is an integer.
All of the orbits discussed in Ref.~\cite{r1} are $\cP\cT$ symmetric, and it was
firmly believed that all closed periodic orbits are $\cP\cT$ symmetric. (We will
see that this is not so, and that non-$\cP\cT$-symmetric orbits are crucial in
understanding the observed rapid variation in the periods of the complex orbits
as $\epsilon$ varies slowly.)

\begin{figure*}[t!]
\vspace{5.25in}
\includegraphics{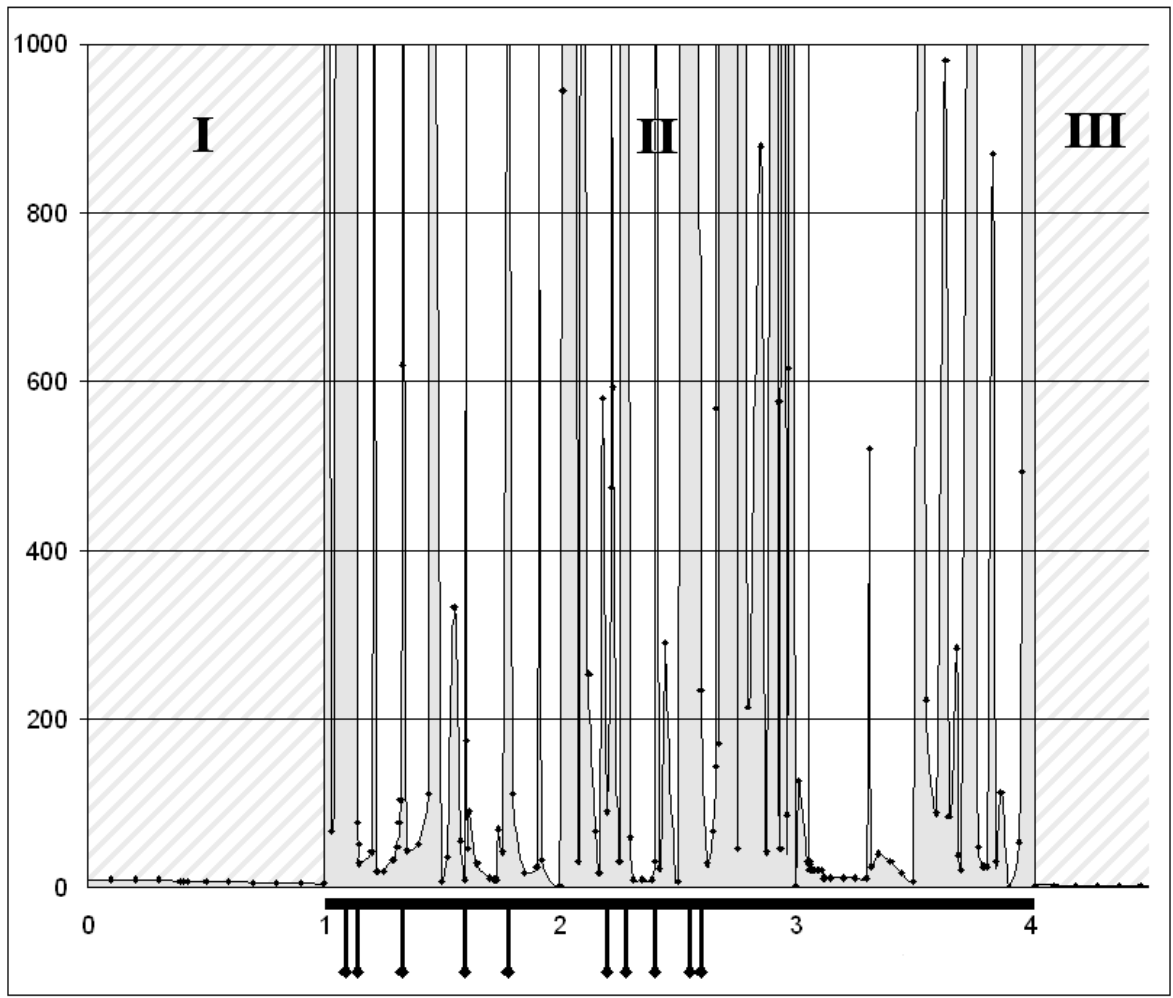}
\caption{Period of a classical trajectory beginning at the $N=1$ turning point
in the complex-$x$ plane. The period is plotted as a function of $\epsilon$. The
period decreases smoothly for $0\leq\epsilon<1$ (Region I). However, when $1\leq
\epsilon\leq4$ (Region II), the period becomes a rapidly varying and noisy
function of $\epsilon$. For $\epsilon>4$ (Region III) the period is once again a
smoothly decaying function of $\epsilon$. Region II contains short subintervals
where the period is a small and smoothly varying function of $\epsilon$. At the
edges of these subintervals the period suddenly becomes extremely long. Detailed
numerical analysis shows that the edges of the subintervals lie at special
rational values of $\epsilon$. Some of these special rational values of
$\epsilon$ are indicated by vertical line segments that cross the horizontal
axis. At these rational values the orbit does not reach the $N=-2$ turning point
and the $\cP\cT$ symmetry of the classical orbit is spontaneously broken.}
\label{fig2}
\end{figure*}

In Ref.~\cite{r1} many complex trajectories $x(t)$ were examined, some having a
rich topological structure. Some of these trajectories visit many sheets of the
Riemann surface. The classical orbits exhibit fine structure that is exquisitely
sensitive to the value of $\epsilon$. Small variations in $\epsilon$ can cause
huge changes in the topology and in the periods of the closed orbits. Depending
on the value of $\epsilon$, there are orbits having short periods as well as
orbits having long and possibly arbitrarily long periods (see Fig.~\ref{fig2}).

\begin{figure*}[t!]
\vspace{2.85in}
\includegraphics{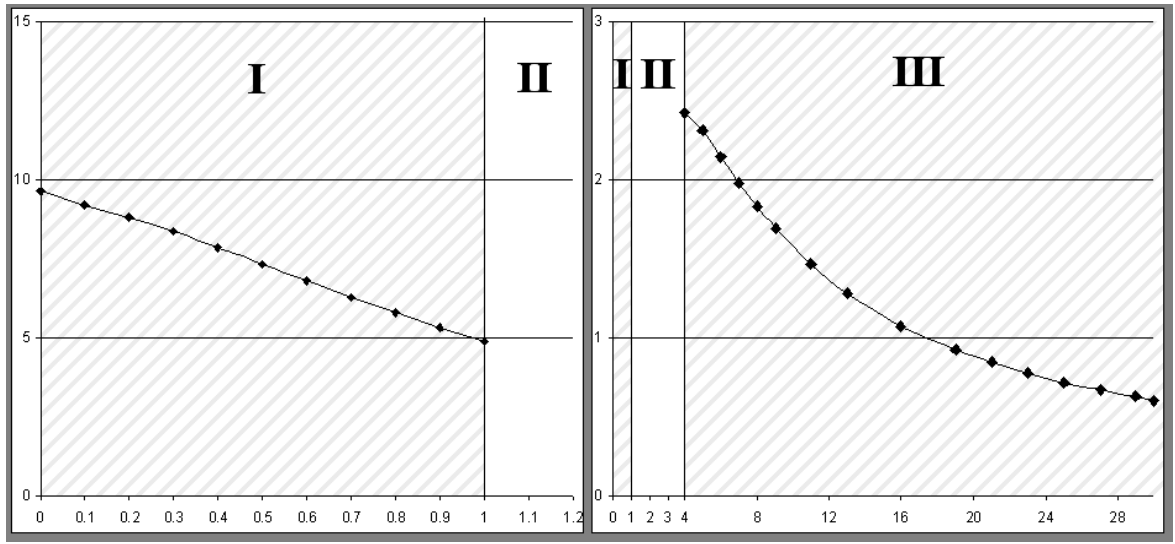}
\caption{Period of a classical trajectory joining the $K=1$ pair of turning
points for $\epsilon$ in Regions I and III. (See Fig.~\ref{fig2}.) The orbits in
these regions are all $\cP\cT$ symmetric.}
\label{fig3}
\end{figure*}

Figure~\ref{fig2} delineates three regions of $\epsilon$ for which an orbit that
begins at the $N=1$ turning point exhibits a specific kind of behavior. When $0
\leq\epsilon\leq1$ (Region I), the period is a smooth decreasing function of
$\epsilon$; when $1<\epsilon\leq4$ (Region II), the period is a rapidly varying
and choppy function of $\epsilon$; when $4<\epsilon$ (Region III), the period is
once again a smooth and decreasing function of $\epsilon$. For some values of
$\epsilon$ in Region II the period is extremely long. Thus, it is difficult to
see the behavior of the period in Regions I and III in Fig.~\ref{fig2}. We have
therefore plotted in Fig.~\ref{fig3} the period for these slowly varying
regions.

\begin{figure*}[t!]
\vspace{5.25in}
\includegraphics{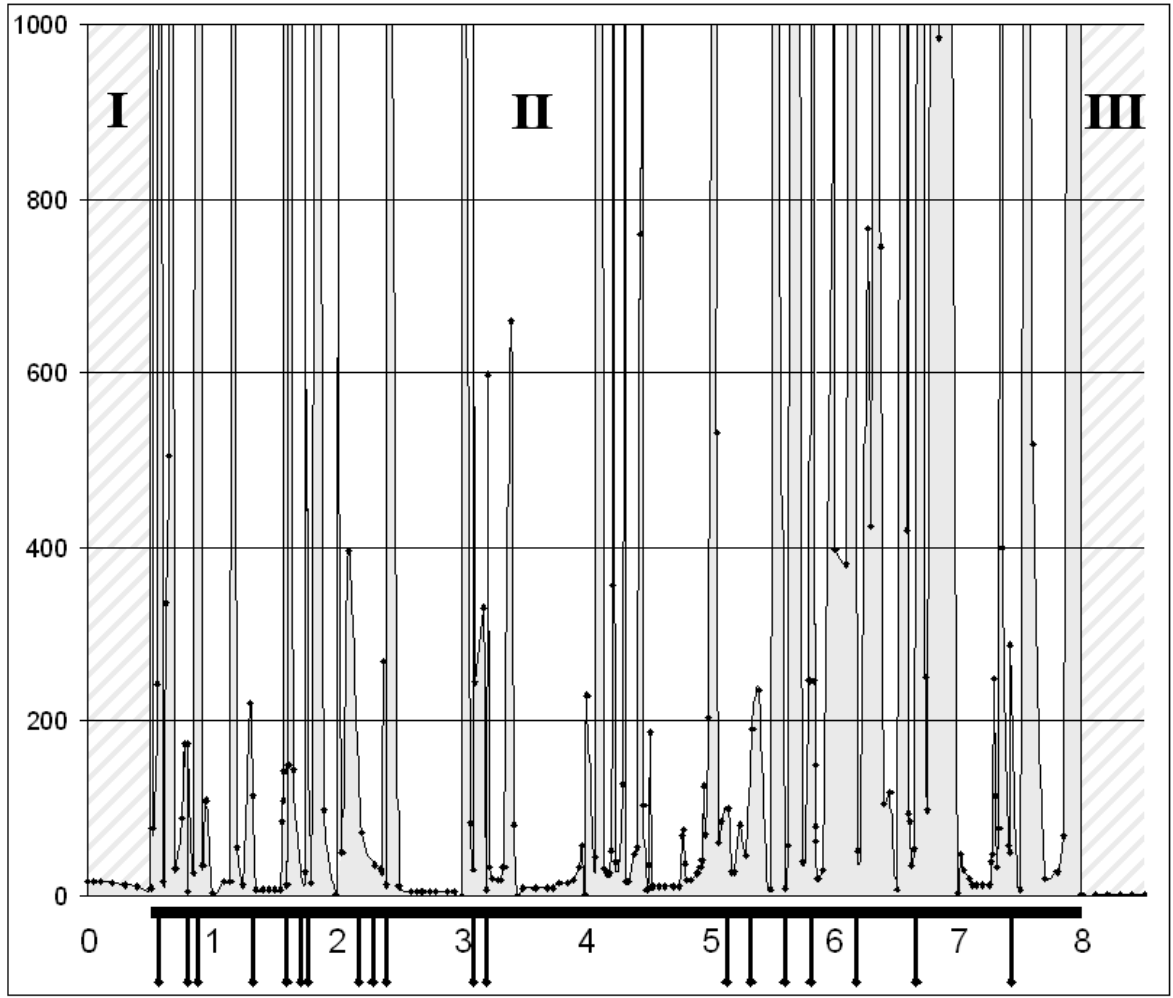}
\caption{Period of a classical trajectory joining (except when $\cP\cT$ symmetry
is broken) the $K=2$ pair of turning points. The period is plotted as a function
of $\epsilon$. As in the $K=1$ case shown in Fig.~\ref{fig2}, there are three
regions. When $0\leq\epsilon\leq\half$ (Region I), the period is a smooth
decreasing function of $\epsilon$; when $\half<\epsilon\leq8$ (Region II), the
period is a rapidly varying and choppy function of $\epsilon$; when $8<\epsilon$
(Region III), the period is again a smooth and decreasing function of
$\epsilon$.}
\label{fig4}
\end{figure*}

\begin{figure*}[t!]
\vspace{2.85in}
\includegraphics{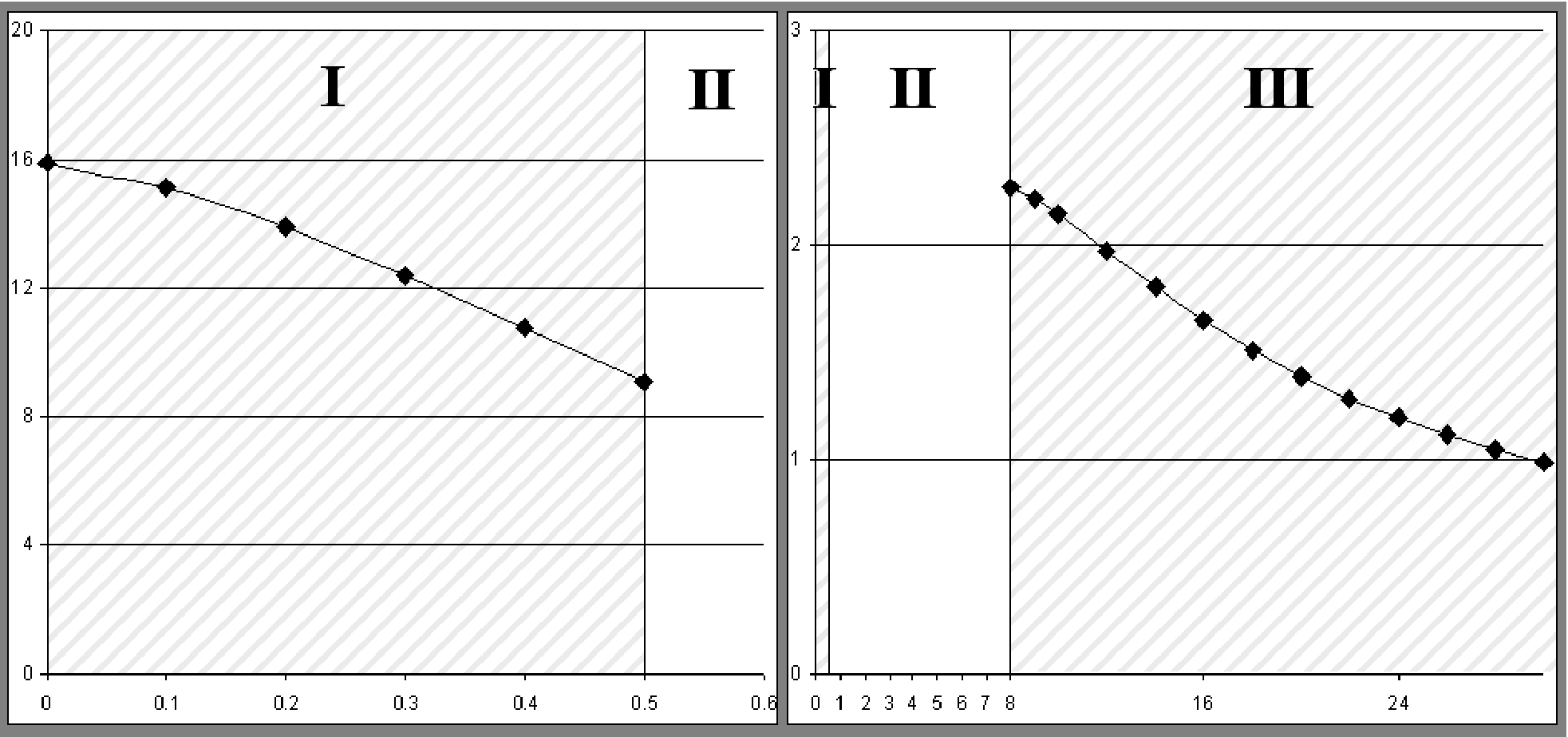}
\caption{Period of a classical trajectory joining the $K=2$ pair of turning
points in the complex-$x$ plane for $\epsilon$ in Regions I and III. (See
Fig.~\ref{fig4}.)}
\label{fig5}
\end{figure*}

For a trajectory beginning at the $N=2$ turning point, the period as a function
of $\epsilon$ again exhibits these three types of behaviors. The period
decreases smoothly for $0\leq\epsilon<\half$ (Region I). When $\half\leq\epsilon
\leq8$ (Region II), the period becomes a rapidly varying and noisy function of
$\epsilon$. When $\epsilon>8$ (Region III) the period is once again a smoothly
decaying function of $\epsilon$. These behaviors are shown in Fig.~\ref{fig4}.
Again, because it is difficult to see the dependence of the period in Regions I
and III when Region II is included, we display the period for $K=2$ for
$\epsilon$ in Regions I and III in Fig.~\ref{fig5}. The trajectory terminates
at the $N=-3$ turning point except when $\cP\cT$ symmetry is spontaneously
broken. Broken-symmetry orbits occur only at isolated points in Region II.

\begin{figure*}[t!]
\vspace{4.85in}
\includegraphics{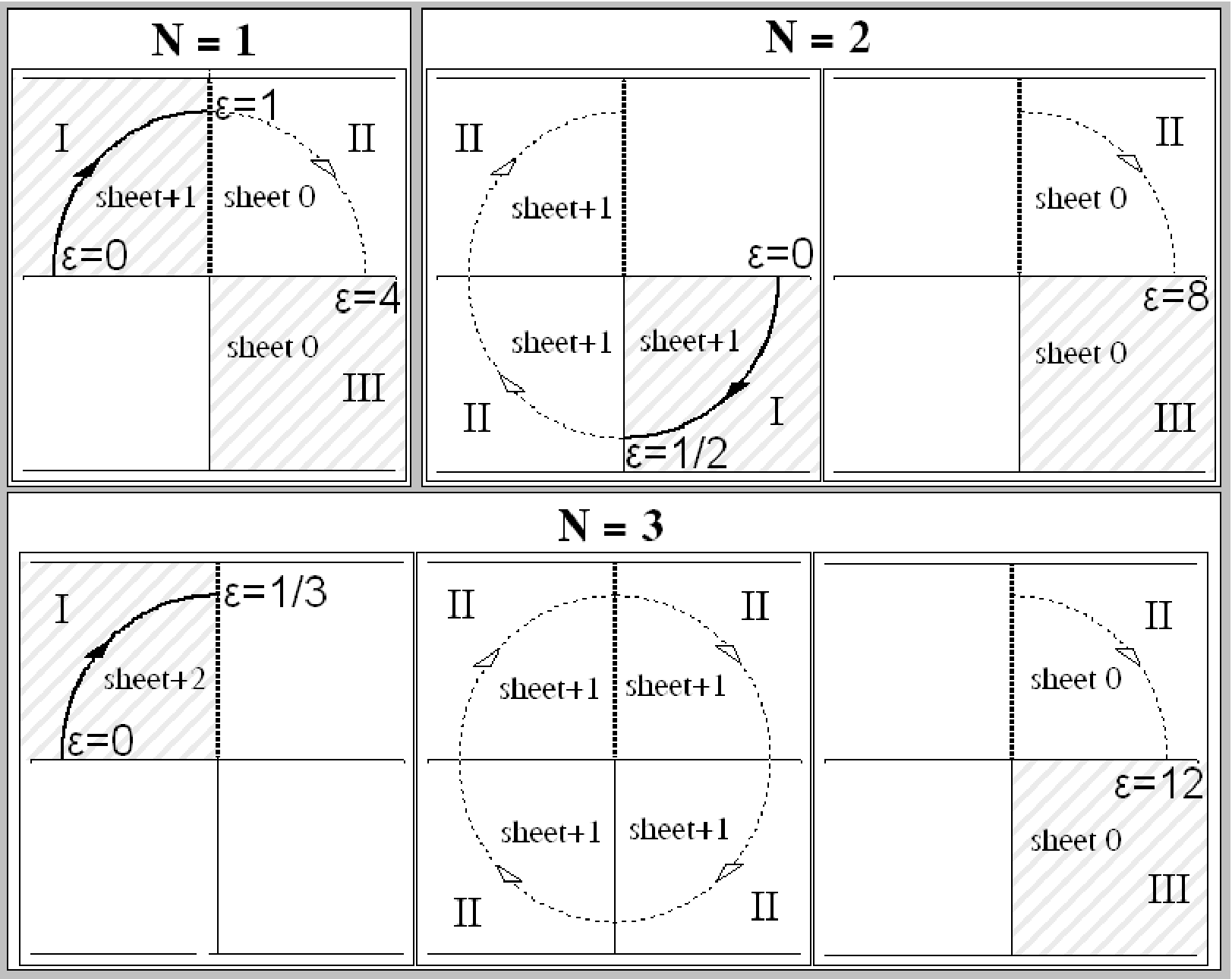}
\caption{Locations of the turning points in the complex-$x$ plane as $\epsilon$
increases from $0$ to $\infty$. At $\epsilon=0$ the turning points lie on the
real-$x$ axis. As $\epsilon$ increases through Region I the turning points
rotate through an angle of $\frac{\pi}{2}$ and end up on the imaginary-$x$ axis.
As $\epsilon$ increases through Region II, the turning points rotate around
the origin and wind up once again on the real axis. Finally, as $\epsilon$
passes through Region III, the turning points again rotate through an angle of
$\frac{\pi}{2}$ and finish up on the negative imaginary-$x$ axis.}
\label{fig6}
\end{figure*}

Figures \ref{fig2} -- \ref{fig5} illustrate a general pattern that holds for all
$K$. For classical orbits that oscillate between the $K$th pair of turning
points, there are always three regions. The domain of Region I is $0\leq\epsilon
\leq\frac{1}{K}$, the domain of Region II is $\frac{1}{K}<\epsilon<4K$, and the
domain of Region III is $4K<\epsilon$. As $\epsilon$ varies, the turning points
move in a characteristic fashion for each of these three regions (see
Fig.~\ref{fig6}). When $\epsilon=0$, the turning points lie on the real axis. As
$\epsilon$ increases, the turning points rotate into the complex-$x$ plane. Just
as $\epsilon$ reaches the upper edge of Region I, the turning points rotate
through an angle of $\frac{\pi}{2}$ and now lie on the imaginary axis. As
$\epsilon$ continues to increase, the turning points continue to rotate around
$x=0$, and may encircle the origin many times. Just as $\epsilon$ reaches the
upper boundary of Region II, the turning points again lie on the real axis.
Finally, as $\epsilon$ moves from the lower edge to the upper edge of Region III
($\epsilon=\infty$), the turning points again rotate through an angle of
$\frac{\pi}{2}$ and lie on the negative imaginary-$x$ axis.

In general, the period of any classical orbit depends on the specific pairs of
turning points that are enclosed by the orbit and on the number of times that
the orbit encircles each pair. As explained in Ref.~\cite{r1}, any orbit can be
deformed to a much simpler orbit of exactly the same period. This simpler orbit
connects two turning points and oscillates between them rather than encircling
them. For the elementary case of orbits that enclose only the $K=0$ pair of
turning points, the formula for the period of the closed orbit is
\begin{eqnarray}
T(\epsilon)=2\sqrt{\pi}\frac{\Gamma\left(\frac{3+\epsilon}{2+\epsilon}\right)}{
\Gamma\left(\frac{4+\epsilon}{r4+2\epsilon}\right)}\cos\left(\frac{\epsilon\pi}{
4+2\epsilon}\right).
\label{e4}
\end{eqnarray}
The derivation of (\ref{e4}) is straightforward. The period $T$ is given by a
closed contour integral along the trajectory in the complex-$x$ plane. This
trajectory encloses the square-root branch cut that joins the $K=0$ pair of
turning points. This contour can be deformed into a pair of rays that run from
one turning point to the origin and then from the origin to the other turning
point. The integral along each ray is easily evaluated as a beta function, which
is then written in terms of gamma functions. Equation (\ref{e4}) is valid for
all $\epsilon\geq0$.

When the classical orbit encloses more than just the $K=0$ pair of turning
points, the formula for the period of the orbit becomes more complicated
\cite{r1}. In general, there are contributions to the period integral from many 
enclosed pairs of turning points. We label each such pair by the integer $j$.
The general formula for the period of the topological class of classical orbits
whose central orbit terminates on the $K$th pair of turning points is
\begin{eqnarray}
T_K(\epsilon)=2\sqrt{\pi}\frac{\Gamma\left(\frac{3+\epsilon}{2+\epsilon}\right)}
{\Gamma\left(\frac{4+\epsilon}{4+2\epsilon}\right)}\sum_{j=0}^{\infty}a_j(K,
\epsilon)\left|\cos\left(\frac{(2j+1)\epsilon\pi}{4+2\epsilon}\right)\right|.
\label{e6}
\end{eqnarray}
In this formula the cosines originate from the angular positions of the turning
points in (\ref{e3}). The coefficients $a_j(K,\epsilon)$ are all nonnegative
integers. The $j$th coefficient is nonzero only if the classical path encloses
the $j$th pair of turning points. Each coefficient is an {\it even\/} integer
except for the $j=K$ coefficient, which is an odd integer. The coefficients $a_j
(K,\epsilon)$ satisfy
\begin{eqnarray}
\sum_{j=0}^{\infty}a_j(K,\epsilon)=k,
\label{e7}
\end{eqnarray}
where $k$ is the number of times that the central classical path crosses the
imaginary axis. Equation (\ref{e7}) truncates the summation in (\ref{e6}) so
that it contains a finite number of terms.

As we can see in Figs.~\ref{fig2} and \ref{fig4}, for an orbit that oscillates
between the $K$th pair of turning points ($K>0$) the classical behavior
undergoes abrupt transitions as $\epsilon$ is varied smoothly in Region II. In
Region II there are narrow patches in which the period of the orbit is rapidly
varying which are sandwiched between small regions of quiet stability. At the
boundaries of the slowly varying and rapidly varying regions there are 
transitions in the topologies and periods of the classical orbits. We can
understand from (\ref{e6}) how there can be rapid variations in the period of
the orbit. The summation in (\ref{e6}) can vary rapidly as a function of
$\epsilon$ because small changes in $\epsilon$ can cause fluctuations in the
topology of the orbit. If the orbit suddenly encloses many more pairs of turning
points, the value of the period may fluctuate wildly. [Note that abrupt changes
in the periods of the orbits cannot occur for the trajectories joining the $K=0$
pairs of turning points because $T(\epsilon)$ in (\ref{e4}) is a smoothly
decreasing function for all $\epsilon\geq0$.]

\section{Classical Orbits Having Broken $\cP\cT$ Symmetry}
\label{s3}

We now demonstrate that the abrupt changes in the topology and the periods of
the orbits that we observe for $\epsilon$ in Region II are associated with the
appearance of orbits having spontaneously broken $\cP\cT$ symmetry. In Region II
there are short patches where the period is relatively small and is a slowly
varying function of $\epsilon$. These patches are bounded by special values of
$\epsilon$ for which the period of the orbit suddenly becomes extremely long.
From our numerical studies of the orbits connecting the $K$th pair of turning
points, we believe that there are only a finite number of these special values
of $\epsilon$ and that these values of $\epsilon$ are always {\it rational}.
Furthermore, we have discovered that at these special rational values of
$\epsilon$, the closed orbits are {\it not} $\cP\cT$-symmetric and we say that
such orbits exhibit {\it spontaneously broken} $\cP\cT$ symmetry. Some special
values of $\epsilon$ at which spontaneously broken $\cP\cT$-symmetric orbits
occur are indicated in Figs.~\ref{fig2} and \ref{fig4} by short vertical lines
below the horizontal axis. These special values of $\epsilon$ always have the 
form $\frac{p}{q}$, where $p$ is a multiple of 4 and $q$ is odd.

\begin{figure*}[t!]
\vspace{2.60in}
\includegraphics{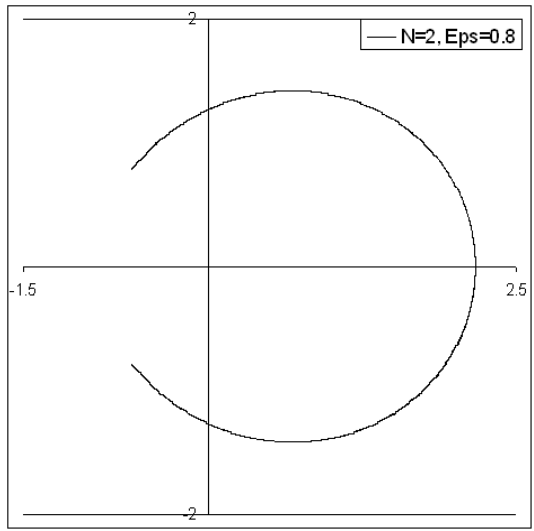}
\caption{A horseshoe-shaped non-$\cP\cT$-symmetric orbit. This orbit is not
symmetric with respect to the imaginary axis but it is symmetric with respect
to the real axis. The orbit terminates at a complex-conjugate pair of turning
points. For this orbit, $\epsilon=\frac{4}{5}$.}
\label{fig7}
\end{figure*}

Figure~\ref{fig7} displays an orbit having spontaneously broken $\cP\cT$
symmetry. This orbit occurs when $\epsilon=\frac{4}{5}$. The orbit starts at the
$N=2$ turning point, but it never reaches the $\cP\cT$-symmetric turning point
$N=-3$. Rather, the orbit terminates when it runs into and is reflected back
from the complex conjugate $N=4$ turning point [see (\ref{e3})]. Thus, a
broken-$\cP\cT$-symmetric orbit is a failed $\cP\cT$-symmetric orbit. The period
of the orbit is short ($T=4.63$). This orbit is not $\cP\cT$ (left-right)
symmetric but it does possess complex-conjugate (up-down) symmetry. In general,
for a non-$\cP\cT$-symmetric orbit to exist, it must join or encircle a pair of
complex-conjugate turning points.

\begin{figure*}[ht!]
\vspace{2.95in}
\includegraphics{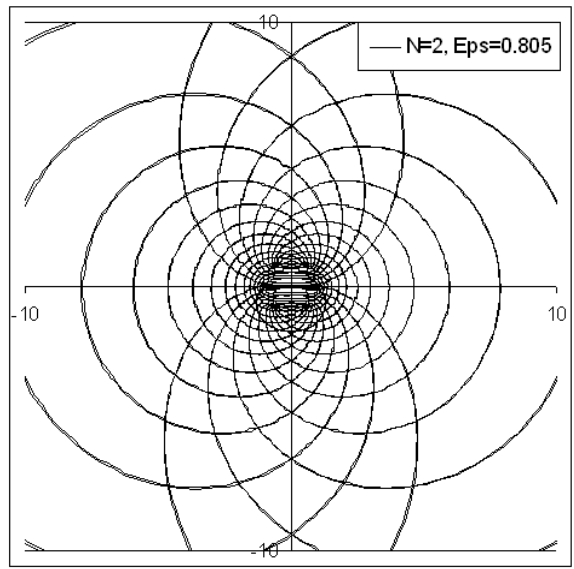}
\caption{$\cP\cT$-symmetric orbit for $\epsilon=0.805$. This orbit connects
the $K=2$ pair of turning points.}
\label{fig8}
\end{figure*}

If we change $\epsilon$ slightly, $\cP\cT$ symmetry is restored and one can only
find orbits that are $\cP\cT$ symmetric. For example, if we take $\epsilon=
0.805$, we obtain the complicated orbit in Fig.~\ref{fig8}. The period of this
orbit is large ($T=173.36$).

\begin{figure*}[t!]
\vspace{2.65in}
\includegraphics{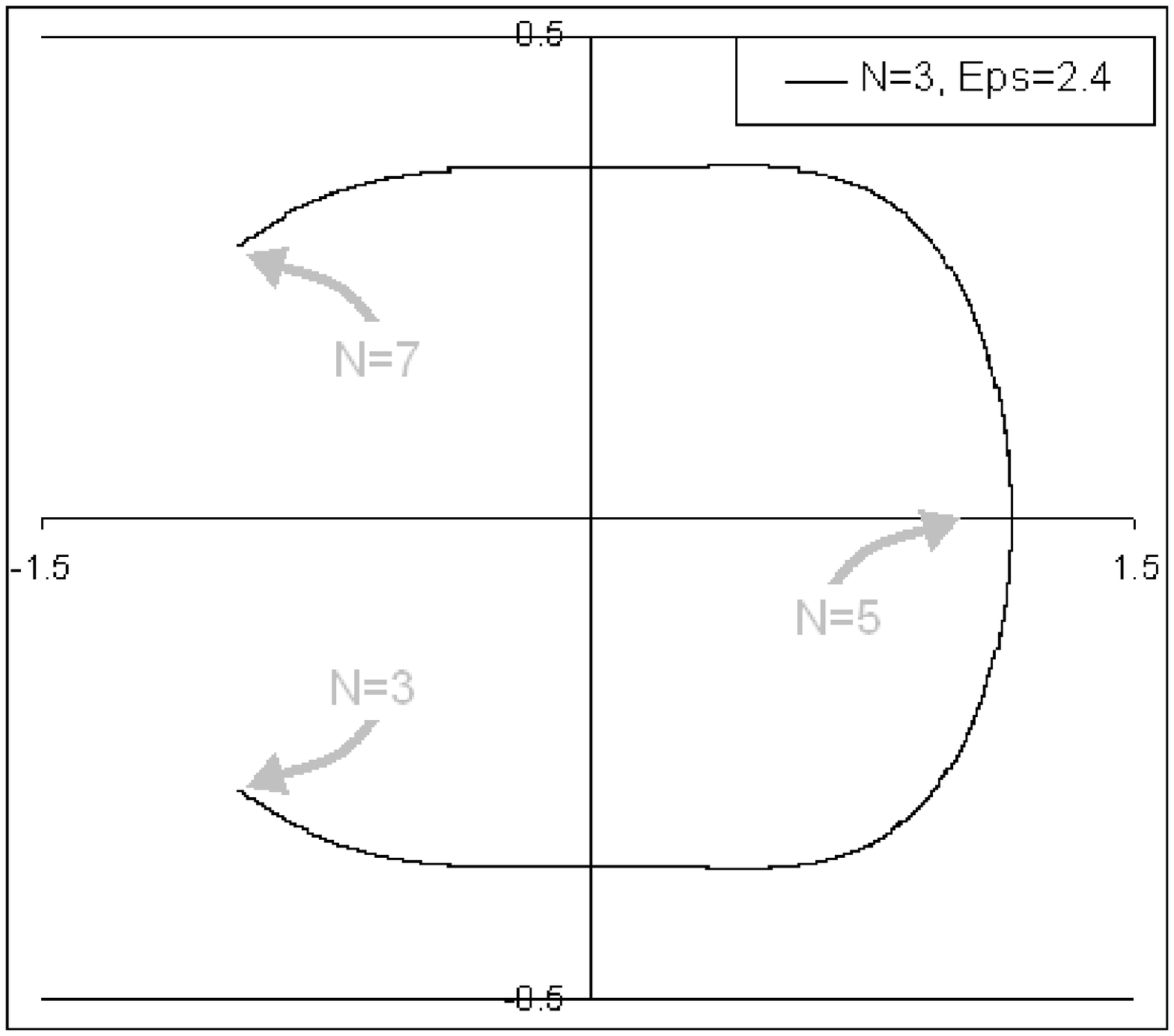}
\caption{A non-$\cP\cT$-symmetric horseshoe-shaped orbit for $\epsilon=\frac{12}
{5}$. This orbit begins at the $N=3$ turning point and terminates at the $N=7$
turning point, so that it fails to reach the $\cP\cT$-symmetric turning point at
$N=-4$. These turning points are shown in Fig.~\ref{fig1}. The period of this
orbit is $T=5.04$.}
\label{fig9}
\end{figure*}

\begin{figure*}[t!]
\vspace{2.75in}
\includegraphics{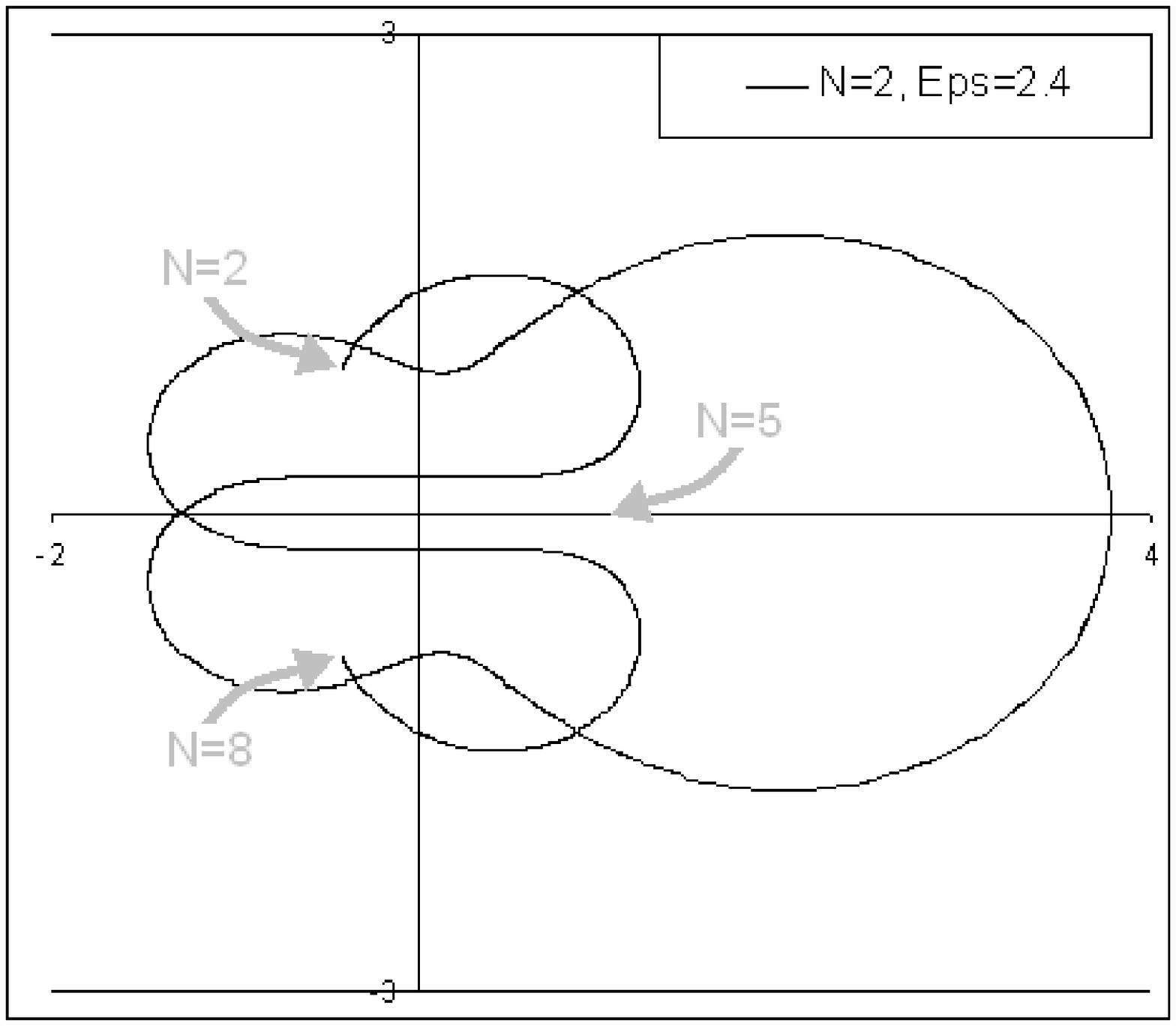}
\caption{A non-$\cP\cT$-symmetric orbit for $\epsilon=\frac{12}{5}$. This orbit,
which is more complicated than that shown in Fig.~\ref{fig9}, begins at the $N=
2$ turning point and ends at the $N=8$ turning point before it can reach the
$\cP\cT$-symmetric turning point at $N=-3$. These turning points are shown in
Fig.~\ref{fig1}. The period of this orbit is $T=12.90$.}
\label{fig10}
\end{figure*}

It is possible to have more than one kind of broken-$\cP\cT$-symmetric orbit for
a given rational value of $\epsilon$. For example, in Figs.~\ref{fig9} and
\ref{fig10} we display two different non-$\cP\cT$-symmetric closed orbits for
$\epsilon=\frac{12}{5}$. In Fig.~\ref{fig9} the horseshoe-shaped orbit
terminates on the $N=3$ and $N=7$ turning points. The period of this orbit is 
($T=5.04$). The more complicated orbit shown in Fig.~10 terminates on the $N=2$
and $N=8$ turning points. The period of this orbit is $T=12.90$. All of these
turning points are shown in Fig.~\ref{fig1} \footnote{In Figs.~\ref{fig9} and
\ref{fig10} the $N=5$ turning point is shown. We have found that for all orbits
having a broken $\cP\cT$ symmetry there exists a special turning point on the
real-$z$ axis. This turning point is symmetric under complex conjugation. A
classical particle that is released from this turning point falls into the
origin, where it stops when it encounters the branch point.}.

\begin{figure*}[t!]
\vspace{3.30in}
\includegraphics{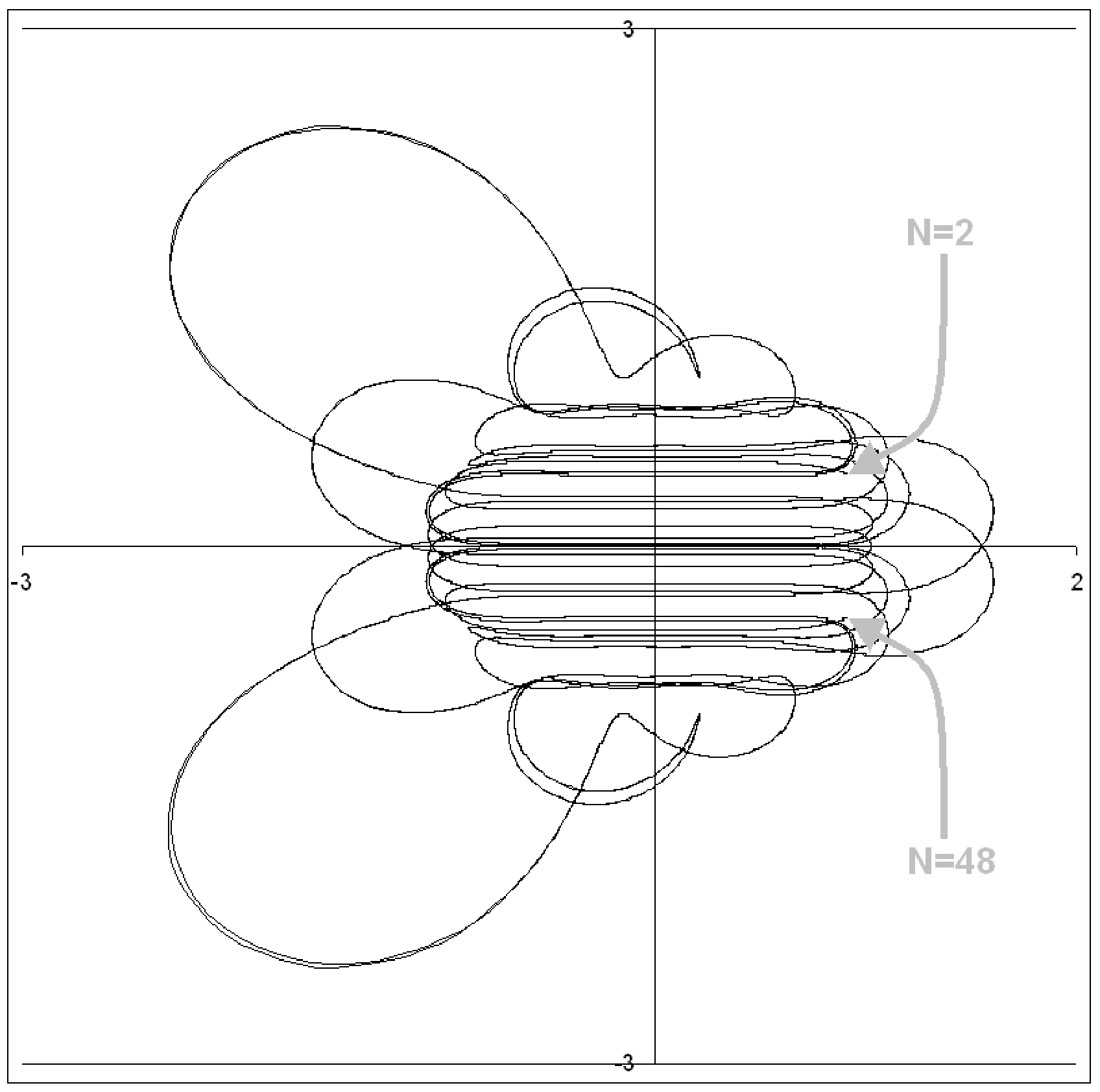}
\caption{A broken-$\cP\cT$-symmetric orbit connecting the $N=2$ and $N=48$
turning points for $\epsilon=\frac{76}{13}$. The period of this orbit is $T=
78.36$.}
\label{fig11}
\end{figure*}

A non-$\cP\cT$-symmetric orbit that is even more complicated than that in
Fig.~\ref{fig10} is shown in Fig.~\ref{fig11}. This orbit connects the $N=2$ and
$N=48$ turning points for $\epsilon=\frac{76}{13}$.

\begin{figure*}[t!]
\vspace{3.25in}
\includegraphics{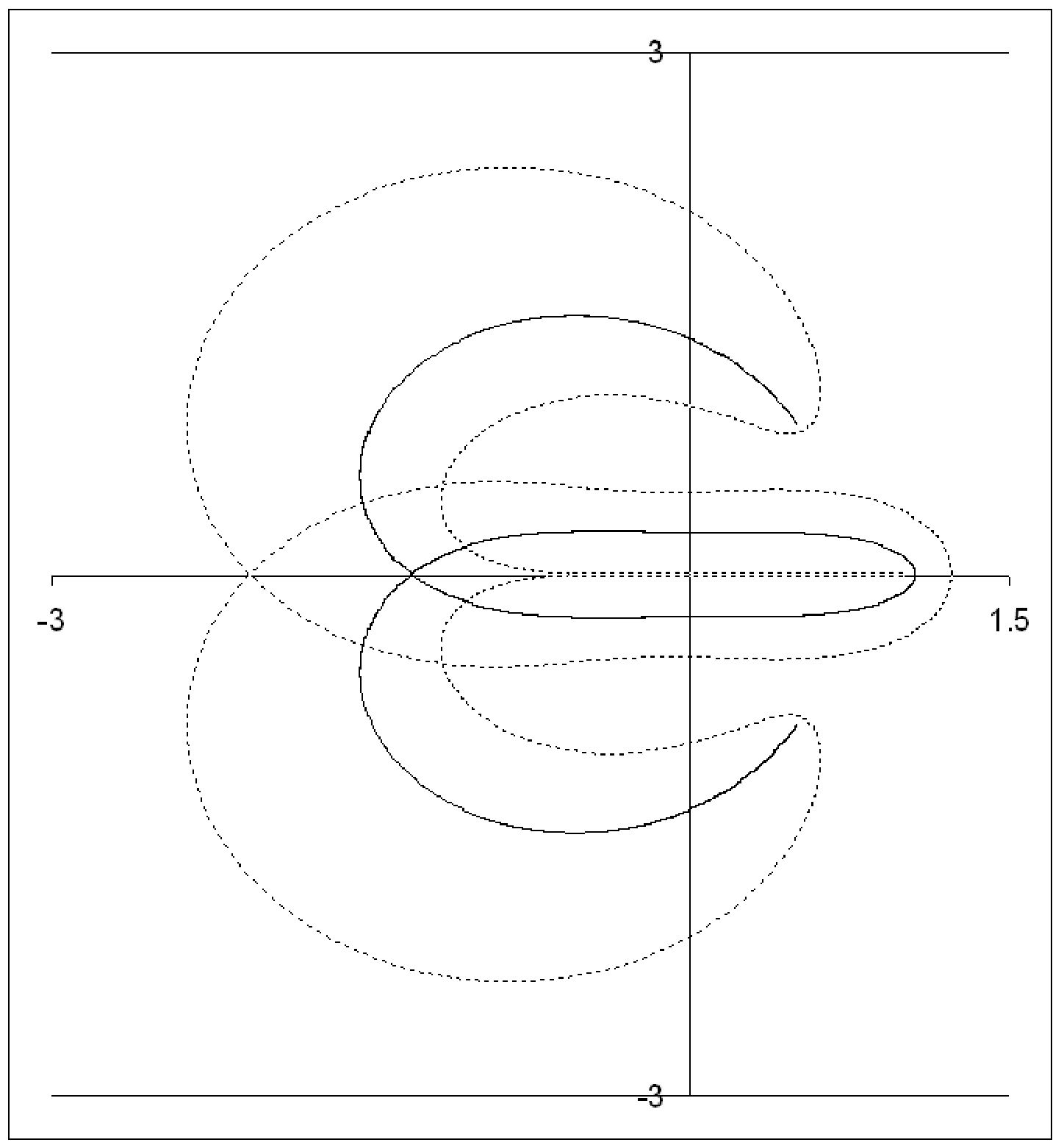}
\caption{Two non-$\cP\cT$-symmetric orbits of the same period $T=9.07$. The
value of $\epsilon$ is $\frac{8}{5}$. The solid-line orbit terminates at the $N=
1$ and $N=7$ turning points, while the dotted-line orbit encircles these turning
points.}
\label{fig12}
\end{figure*}

Non-$\cP\cT$-symmetric orbits can encircle a complex-conjugate pair of turning
points as well as terminating at them. In Fig.~\ref{fig12} two non-$\cP
\cT$-symmetric orbits are shown for $\epsilon=\frac{8}{5}$. One orbit joins the
$N=1$ and $N=7$ complex-conjugate pair of turning points. The other orbit
encircles these turning points. Both orbits have the same period of $T=9.07$.

\begin{figure*}[t!]
\vspace{3.65in}
\includegraphics{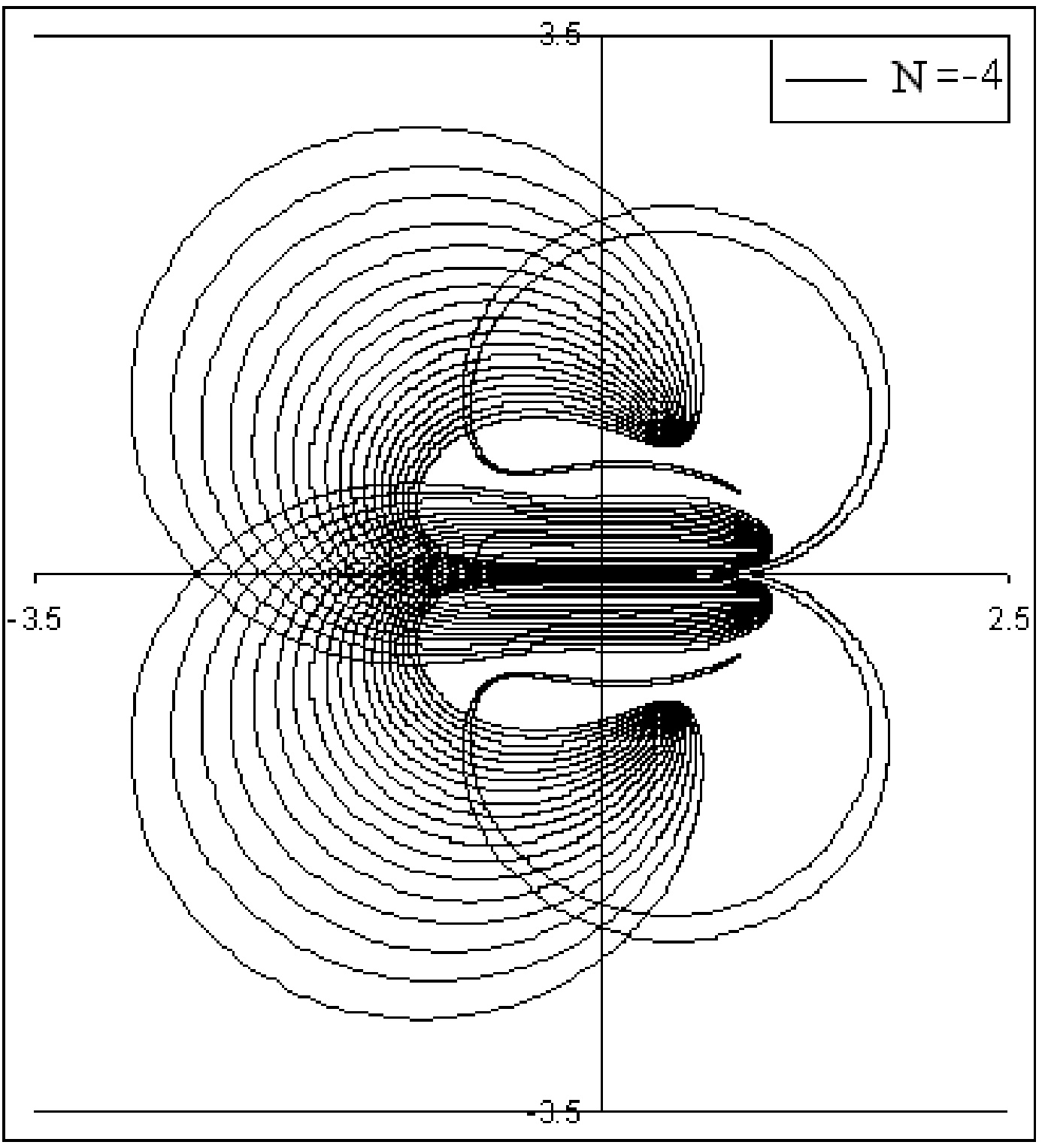}
\caption{Non-$\cP\cT$-symmetric orbit for $\epsilon=\frac{16}{9}$. This
topologically complicated orbit originates at the $N=-4$ turning point but
does not reach the $\cP\cT$-symmetric $N=3$ turning point. Instead, it is
reflected back at the complex-conjugate $N=-14$ turning point. The period of
this orbit is $T=186.14$).}
\label{fig13}
\end{figure*}

\begin{figure*}[t!]
\vspace{3.7in}
\includegraphics{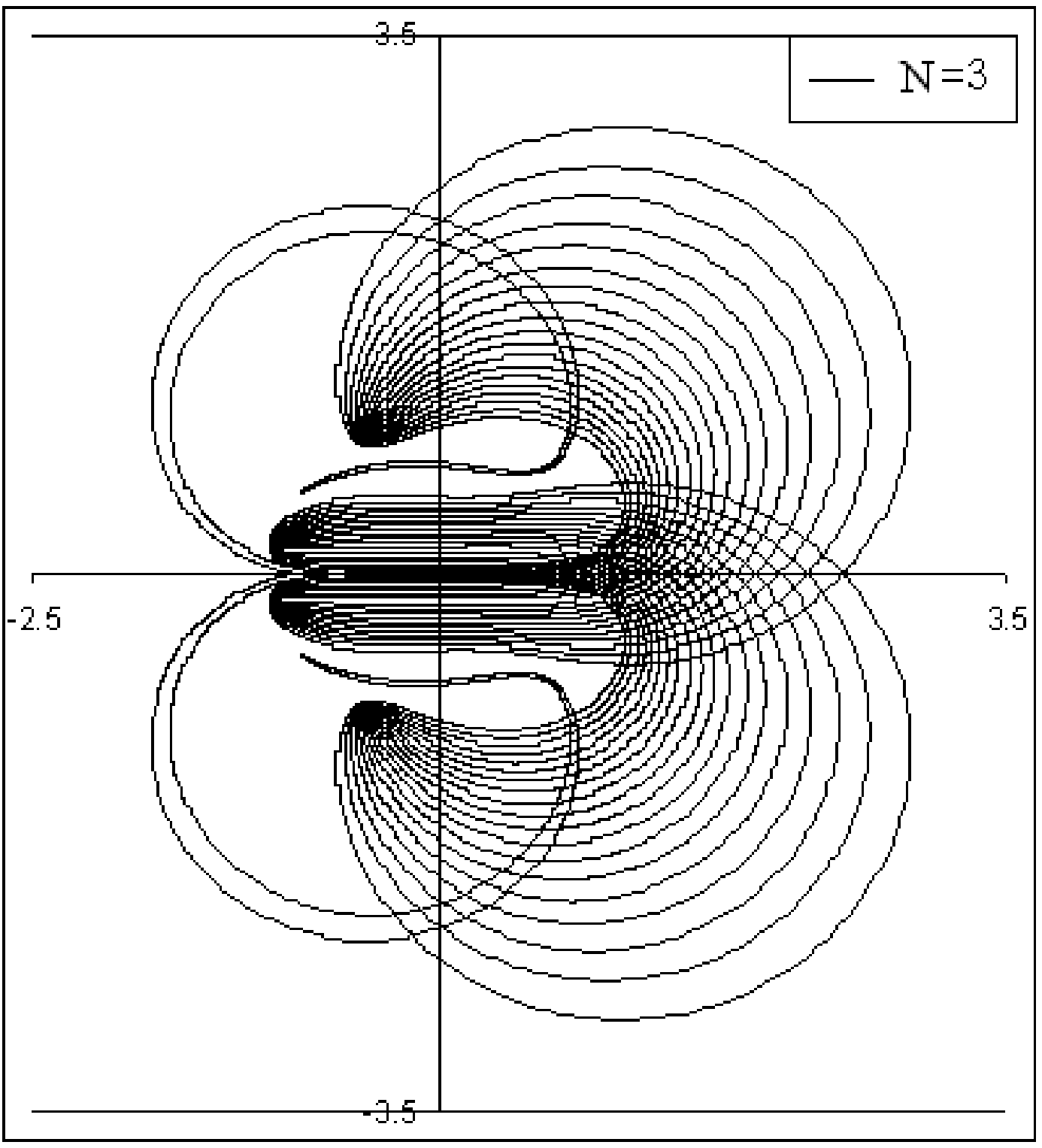}
\caption{$\cP\cT$-symmetric reflection of the orbit in Fig.~\ref{fig13}. This
orbit originates at the $N=3$ turning point but is reflected back by the $N=13$
turning point.}
\label{fig14}
\end{figure*}

Broken-$\cP\cT$-symmetric orbits can have an elaborate topology. For example, at
$\epsilon=\frac{16}{9}$ we find a non-$\cP\cT$-symmetric orbit whose topology is
even more complicated than that of the orbit shown in Fig.~\ref{fig11}. This
orbit, which is shown in Fig.~\ref{fig13}, is a failed $K=3$ $\cP\cT$-symmetric
orbit. It originates at the $N=-4$ turning point, but it never reaches the $\cP
\cT$-symmetric $N=3$ turning point. This is because it is reflected back by the
complex-conjugate $N=-14$ turning point. Figure \ref{fig14} shows the $\cP
\cT$-symmetric companion of the orbit in Fig.~\ref{fig13}. This orbit begins at
the $N=3$ turning point, but is reflected back by the $N=13$ turning point,
which is the $\cP\cT$ counterpart of the $N=-14$ turning point.

\begin{figure*}[t!]
\vspace{3.6in}
\includegraphics{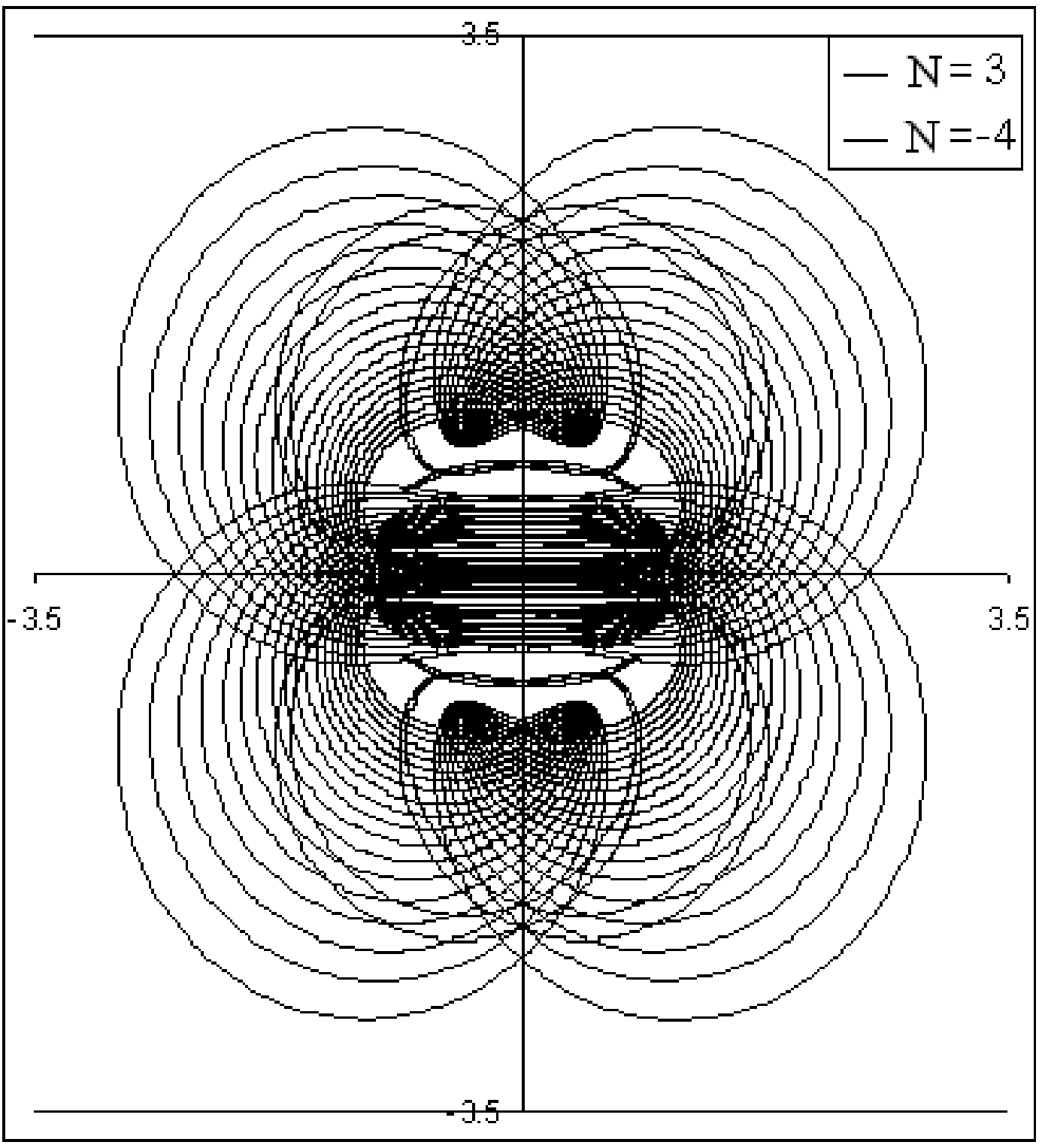}
\caption{Superposition of the orbits in Figs.~\ref{fig13} and \ref{fig14}.
The $\cP\cT$ (left-right) symmetry is exact.}
\label{fig15}
\end{figure*}

\begin{figure*}[t!]
\vspace{4.7in}
\includegraphics{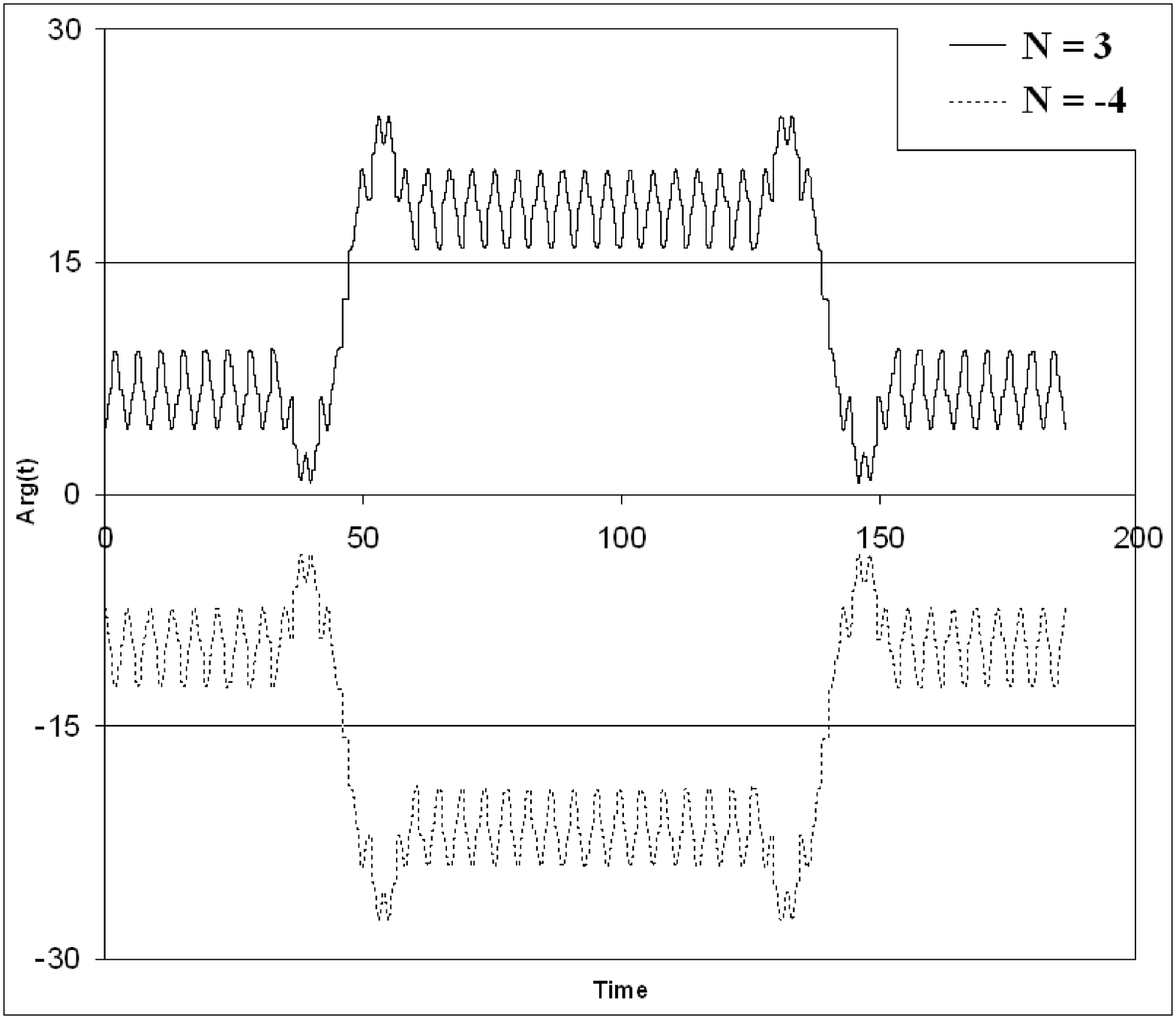}
\caption{The complex argument of $x(t)$ plotted as a function of $t$ for the
non-$\cP\cT$-symmetric orbits in Figs.~\ref{fig13} and \ref{fig14}. The two
plots verify that the two orbits are $\cP\cT$-symmetric reflections of one
another.}
\label{fig16}
\end{figure*}

To show that the orbits in Figs.~\ref{fig13} and \ref{fig14} are $\cP\cT$
reflections, we have plotted both orbits in Fig.~\ref{fig15}. The left-right
symmetry is manifest. A definitive demonstration of the symmetry can be given by
plotting the complex argument of $x(t)$ as a function of time $t$ for each
orbit. This is done in Fig.~\ref{fig16}.

\begin{figure*}[t!]
\vspace{6.25in}
\includegraphics{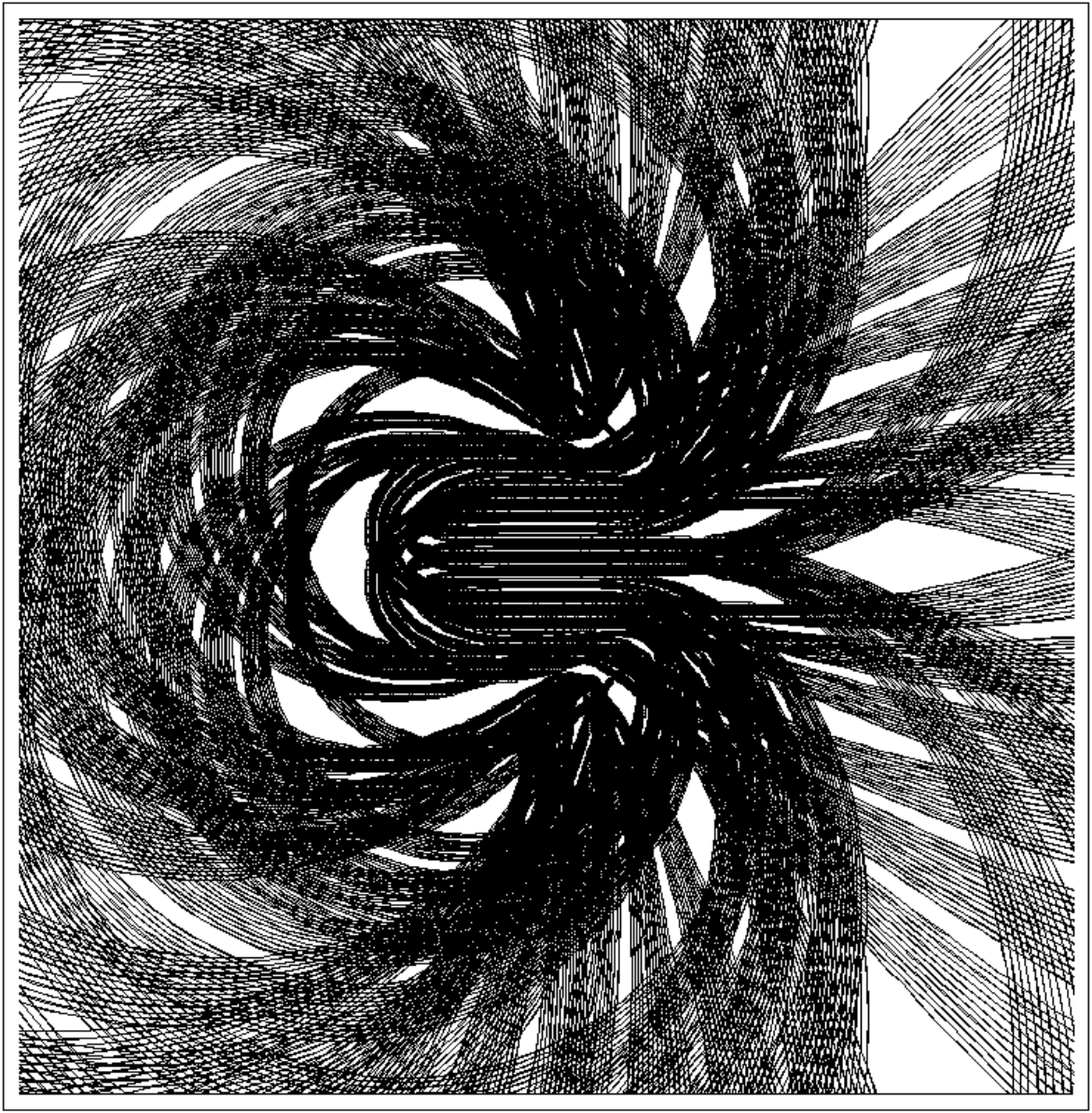}
\caption{A non-$\cP\cT$-symmetric orbit having an extremely elaborate topology.
The value of $\epsilon$ for this orbit is $\frac{16}{15}$. The orbit originates
at the $N=1$ turning point and terminates at the complex-conjugate $N=21$
turning point. This orbit visits eight of the fifteen sheets of the Riemann
surface and travels a great distance from the origin. Only the $-2<{\rm Re}\,x<
2$, $-2<{\rm Im}\,x<2$ portion of the complex-$x$ plane is shown.}
\label{fig17}
\end{figure*}

Finally, we display in Fig.~\ref{fig17} a spontaneously broken-$\cP
\cT$-symmetric orbit that is vastly more complicated than those shown in
Figs.~\ref{fig13} and \ref{fig14}. This orbit begins at the $N=1$ turning point
for $\epsilon=\frac{16}{15}$. It terminates at the complex conjugate $N=21$
turning point rather than at the $\cP\cT$-symmetric $N=-2$ turning point. The
period of this orbit is $T=3393.64$.

\section{Concluding Remarks}
\label{s4}

The work in Ref.~\cite{r1} provides a heuristic explanation of how very
long-period orbits arise. In order for a classical trajectory to travel a great
distance in the complex plane, its path must slip through a forest of turning
points. When the trajectory comes under the influence of a turning point, it
usually executes a huge number of nested U-turns and eventually returns back to
its starting point. However, for some values of $\epsilon$ the complex
trajectory may evade many turning points before it eventually encounters a
turning point that takes control of the particle and flings it back to its
starting point. We speculated in Ref.~\cite{r1} that it may be possible to find
special values of $\epsilon$ for which the classical path manages to avoid and
sneak past all turning points. Such a path would have an infinitely long period.
(We still do no know if such infinite-period orbits exist.)

However, in Ref.~\cite{r1} we could not provide an explanation of why the period
of a closed trajectory, as a function of $\epsilon$, is such a wildly
fluctuating function. We have shown here that for special rational values of
$\epsilon$ the trajectory bumps directly into a turning point that is located 
at a point that is the complex conjugate of the point from which the trajectory
was launched. This turning point reflects the trajectory back to its starting
point and prevents the trajectory from being $\cP\cT$ symmetric. Trajectories
for values of $\epsilon$ near these special rational values have extremely
different topologies and thus have periods that tend to be relatively long. This
explains the noisy plots in Figs.~\ref{fig2} and \ref{fig4}.

Figs.~\ref{fig7} and \ref{fig8} illustrate this phenomenon. The value of
$\epsilon$ for Fig.~\ref{fig7} differs from that in Fig.~\ref{fig8} by only
$0.005$. Nevertheless, the orbits in these two figures exhibit very different
topologies. This is because the trajectory in Fig.~\ref{fig7} is a failed $\cP
\cT$-symmetric orbit, where the trajectory that starts at the $N=2$ turning
point is reflected almost immediately by the complex-conjugate $N=4$ turning
point. When $\epsilon$ is changed by the tiniest amount, the turning points are
slightly displaced. As a result, the trajectory in Fig.~\ref{fig8} manages to
sneak past the $N=4$ turning point and travels a great distance in the complex
plane before bouncing back from the $\cP\cT$-symmetric $N=-3$ turning point.

We do not know whether for each turning point there are a finite or an infinite
number of special rational values of $\epsilon$ for which the classical orbit
has a broken $\cP\cT$ symmetry. It is worth noting that the data used to produce
Figs.~\ref{fig2} and \ref{fig4} is far from exhaustive and that the bars
underneath the horizontal axes are only the {\em known} examples of broken
$\cP\cT$ symmetry. There are surely many more such special rational values of
$\epsilon=\frac{p}{q}$. In this paper we concluded our study at $q=13$, and
this work took an immense amount of computer time!

The study of complex trajectories for classical dynamical systems is a rich new
area in mathematical physics that deserves extensive analytical and numerical
exploration. Already, there has been work done on the complex orbits of the
simple pendulum \cite{r7}, the complex extension of the Korteweg-de Vries
equation \cite{r8,r9}, complex solutions of the Euler equations for rigid-body
rotation \cite{r10}, and the complex version of the kicked rotor \cite{r11}. We
expect that complex analysis will provide a deep insight into the behavior of
dynamical systems.

\begin{acknowledgments}
We thank D.~Hook and S.~McLenahan for programming assistance. CMB is grateful to
the Theoretical Physics Group at Imperial College, London, for its hospitality.
As an Ulam Scholar, CMB receives financial support from the Center for Nonlinear
Studies at the Los Alamos National Laboratory and he is supported in part by a
grant from the U.S. Department of Energy.
\end{acknowledgments}

\end{document}